# QED using the Nilpotent Formalism

Peter Rowlands* and J. P. Cullerne†

*IQ Group and Science Communication Unit, Department of Physics, University of Liverpool, Oliver Lodge Laboratory, Oxford Street, P.O. Box 147, Liverpool, L69 7ZE, UK. e-mail prowl@hep.ph.liv.ac.uk and prowl@csc.liv.uk

†IQ Group, Department of Computer Science, University of Liverpool, Chadwick Laboratory, Peach Street, Liverpool, L69 7ZF, UK.

*Abstract.* The nilpotent formalism for the Dirac equation, outlined in previous papers, is applied to QED. It is shown that what is usually described as 'renormalization' is effectively a statement of the fact that the nilpotent formulation is automatically second quantized and constrains the quantum field into producing finite values for fundamental physical quantities.

## 1 Introduction

In earlier papers[1-3] we have developed a formalism for the Dirac equation, based on a nilpotent algebra, which has been applied in such cases as the hydrogen atom and the strong interacting potential. The idea has been to generate structures which explain various aspects of particle physics in a more fundamental way. Certain aspects of this algebra have also suggested that it might lead to a more fundamental understanding of the existence of apparent infinities in the QED formalism and their removal by the method of renormalization. It seems possible that a formalism which doesn't appear to require normalization, might not require renormalization either.[4] Also certain aspects of the algebra indicate that it has a built-in second quantization and a natural supersymmetry without requiring extra supersymmetric particles, which suggests that infinite sums must add up to finite values and that any necessary cancellations might be automatic.[2] Our aim, therefore, has been to develop our formalism to incorporate the existing techniques of QED, and to discover if there are any aspects of the formalism which help either to explain or to remove apparent anomalies in the existing structures. A brief discussion of significant aspects of the nilpotent algebra will be followed by an outline derivation of those aspects of QED which are relevant to our discussion of the basis of the apparent divergences in the theory and their resolution through renormalization.

## 2 The nilpotent formulation of relativistic quantum mechanics

The nilpotent formulation requires the use of the 32-part algebra generated by combining the 4-vector units ($i$, **i**, **j**, **k**) with the quaternion units (1, $i$, $j$, $k$), where the two sets of terms correspond to the units of the fundamental parameters space-time and mass-charge. The vector units (**i**, **j**, **k**), however, have the multivariate or quaternion-like properties of Pauli matrices. In effect, this means defining a 'full product' for two vectors **a** and **b** of the form

$$\mathbf{ab} = \mathbf{a}.\mathbf{b} + i\,\mathbf{a} \times \mathbf{b},$$

and using the following rules for the multiplication of the vector units:

$$\mathbf{i}^2 = \mathbf{j}^2 = \mathbf{k}^2 = 1$$
$$\mathbf{ij} = -\mathbf{ji} = i\mathbf{k}$$
$$\mathbf{jk} = -\mathbf{kj} = i\mathbf{i}$$
$$\mathbf{ki} = -\mathbf{ik} = i\mathbf{j}.$$

To construct the Dirac equation using this algebra, we begin with the classical relativistic energy conservation equation,

$$E^2 - p^2 - m^2 = 0,$$

which we factorize and supply with the exponential term $e^{-i(Et - \mathbf{p}.\mathbf{r})}$, so that

$$(\pm\,\mathbf{k}E \pm i\mathbf{i}\,\mathbf{p} + i\mathbf{j}\,m)\,(\pm\,\mathbf{k}E \pm i\mathbf{i}\,\mathbf{p} + i\mathbf{j}\,m)\,e^{-i(Et - \mathbf{p}.\mathbf{r})} = 0.$$

Then replacing $E$ and $\mathbf{p}$ in the first bracket with the quantum operators, $i\partial/\partial t$ and $-i\nabla$, to give

$$\left(\pm\,i\mathbf{k}\frac{\partial}{\partial t} \pm i\nabla + i\mathbf{j}m\right)(\pm\,\mathbf{k}E \pm i\mathbf{i}\,\mathbf{p} + i\mathbf{j}\,m)\,e^{-i(Et - \mathbf{p}.\mathbf{r})} = 0,$$

we obtain a quantum mechanical equation of the form

$$\left(\pm\,i\mathbf{k}\frac{\partial}{\partial t} \pm i\nabla + i\mathbf{j}m\right)\psi = 0,$$

where the wavefunction (or, in our terminology, the quaternion state vector) for a free fermion becomes

$$\psi = (\pm\,\mathbf{k}E \pm i\mathbf{i}\,\mathbf{p} + i\mathbf{j}\,m)\,e^{-i(Et - \mathbf{p}.\mathbf{r})},$$

and the following correspondences are seen between the conventional gamma matrices and the terms of the nilpotent algebra:

$$\gamma^0 = i\mathbf{k},\ \gamma^1 = i\mathbf{i},\ \gamma^2 = i\mathbf{j},\ \gamma^3 = i\mathbf{k},\ \gamma^5 = i\mathbf{j}.$$

For convenience, we generally arrange the four terms incorporated in $\psi$ in a column (or ket) vector, while the four terms in the differential operator take up equivalent positions in a row (or bra) vector.

The quaternion state vectors in this formulation have remarkable properties. They are nilpotents or square roots of 0, and they have the properties of quantum field operators, creation and annihilation operators, and even supersymmetry operators. Because the differential operators are, in effect, identical to the quaternion state vectors, the theory already has the second quantization required of quantum field theory, and it has been rigorously shown that they are identical to quantum field integrals acting on vacuum.[1] Another important aspect of the nilpotent state vector is that it automatically incorporates both nonlocality (because a superposition of two identical wavefunctions is instantaneously impossible with a nilpotent structure) and time-delayed local action (because nilpotents are square roots of a relativistic equation).

Many important results have already been achieved with this formalism, but, for the purposes of this paper, we will simply outline a few important technical points. Standard normalization techniques may be used if required, though the nilpotent structure usually means they are not usually necessary. (The nilpotents used here have been multiplied by an extra *i* for mathematical convenience, and this would be removed in normalization.) The momentum or angular momentum operator **p**, being multivariate, can be multiplied with itself to give $p^2$, as can the scalar product σ.**p**, and so it is possible to use either **p** or σ.**p** in the quaternion state vector. If fermions can be taken as having the four terms in the state vector in a particular order, say,

$$(kE + ii\,\mathbf{p} + ij\,m)\ ;\ (kE - ii\,\mathbf{p} + ij\,m);\ (-kE + ii\,\mathbf{p} + ij\,m);\ (-kE - ii\,\mathbf{p} + ij\,m)\ ,$$

then the order of terms in the equivalent antifermion (with the same spin) is also determined as:

$$(-kE + ii\,\mathbf{p} + ij\,m)\ ;\ (-kE - ii\,\mathbf{p} + ij\,m);\ (kE + ii\,\mathbf{p} + ij\,m);\ (kE - ii\,\mathbf{p} + ij\,m)\ .$$

To specify either a fermion or antifermion, it will often be convenient just to specify the first of the four terms, with the presence of the other three automatically understood. We will, for example, refer on occasions to

$$(kE + ii\,\mathbf{p} + ij\,m)\ (-kE + ii\,\mathbf{p} + ij\,m)$$

being a scalar on the basis of this understanding.

Reversing the fermion or antifermion spin state requires reversing the signs of **p** in each term. Boson state vectors are constructed from the scalar product of a fermion and an antifermion state vector, arranged as bra and ket (in either order), for example,

$$(kE + ii\,\mathbf{p} + ij\,m)\ (-kE + ii\,\mathbf{p} + ij\,m) + (kE - ii\,\mathbf{p} + ij\,m)\ (-kE - ii\,\mathbf{p} + ij\,m)$$

$$+\ (-kE + ii\,\mathbf{p} + ij\,m)\ (kE + ii\,\mathbf{p} + ij\,m) + (-kE - ii\,\mathbf{p} + ij\,m)\ (kE - ii\,\mathbf{p} + ij\,m)\ ,$$

and are always pure scalars (in this case, $8E^2$ before normalization). The expression, here, gives the general case; many bosons, of course, will have $m = 0$. It is also the expression for a vector boson. A scalar boson would have $(kE + ii\,\mathbf{p} + ij\,m)(-kE - ii\,\mathbf{p} + ij\,m)\ldots$, and cannot have $m = 0$, as $(kE + ii\,\mathbf{p})(-kE - ii\,\mathbf{p})\ldots$ then disappears.

A vacuum operator applied simultaneously to all four solutions is most conveniently represented by a diagonal matrix, premultiplied by a row state vector or postmultiplied by a column state vector. The vacuum wavefunction operator (when applied to a row vector) is always $k \times$ matrix form of state vector $\times$ exponential term:

$$k \begin{pmatrix} kE + ii\mathbf{p} + ijm & 0 & 0 & 0 \\ 0 & kE - ii\mathbf{p} + ijm & 0 & 0 \\ 0 & 0 & -kE + ii\mathbf{p} + ijm & 0 \\ 0 & 0 & 0 & -kE - ii\mathbf{p} + ijm \end{pmatrix} e^{-i(Et - \mathbf{p}\cdot\mathbf{r})}$$

The order is reversed when applied to a column vector. The vacuum operator omits the exponential term. Applying the vacuum operator leaves the form of a fermion or antifermion quaternion state vector unchanged. It appears to be significant that the infinite series

$$(kE + ii\mathbf{p} + ijm)\,k\,(kE + ii\mathbf{p} + ijm)\,k\,(kE + ii\mathbf{p} + ijm)\,k\,(kE + ii\mathbf{p} + ijm)\ldots,$$

which may be taken as representative as the individual terms in a fermion state vector acting on vacuum, is identical to

$$(kE + ii\mathbf{p} + ijm)(-kE + ii\mathbf{p} + ijm)(kE + ii\mathbf{p} + ijm)(-kE + ii\mathbf{p} + ijm)\ldots,$$

which may be taken as the creation by a fermion of an infinite series of alternate boson and fermion structures, presumably in the vacuum in a kind of supersymmetric series, with each particle or particle plus vacuum being its own supersymmetric partner. This suggests the existence of a kind of supersymmetry without a new set of particles.

## 3 A perturbation expansion of the Dirac equation for QED

We begin with the Dirac equation for a fermion in the presence of the electromagnetic potentials $\phi$, $\mathbf{A}$.

$$\left(k\frac{\partial}{\partial t} + i\,i\,\sigma.\nabla + ijm\right)\psi = -e\,(i\,k\phi - i\,\sigma.\mathbf{A})\,\psi$$

and apply a perturbation expansion to $\psi$, so that

$$\psi = \psi_0 + \psi_1 + \psi_2 + \ldots,$$

where $\psi_0 = (kE + ii\sigma.\mathbf{p} + ijm)\, e^{-i(Et - \mathbf{p}.\mathbf{r})}$ is the solution of the unperturbed equation:

$$\left(k\frac{\partial}{\partial t} + ii\,\sigma.\nabla + ijm\right)\psi_0 = 0,$$

and represents zeroth-order coupling, or a free electron of momentum $\mathbf{p}$. Here, of course, as always, we take $(kE + ii\sigma.\mathbf{p} + ijm)$, and the equivalent differential operator, as a 4-component vector incorporating the four solutions $(\pm kE \pm ii\sigma.\mathbf{p} + ijm)$ in a fixed order, say $(kE + ii\sigma.\mathbf{p} + ijm)$, $(-kE + ii\sigma.\mathbf{p} + ijm)$, $(kE - ii\sigma.\mathbf{p} + ijm)$, $(-kE - ii\sigma.\mathbf{p} + ijm)$. This means that $(kE + ii\sigma.\mathbf{p} + ijm)(-kE + ii\sigma.\mathbf{p} + ijm)$ becomes a scalar quantity, while $(kE + ii\sigma.\mathbf{p} + ijm)(kE + ii\sigma.\mathbf{p} + ijm) = 0$.

Using the perturbation expansion, we can write

$$\left(k\frac{\partial}{\partial t} + ii\,\sigma.\nabla + ijm\right)(\psi_0 + \psi_1 + \psi_2 + \ldots) = -e\,(ik\phi - i\,\sigma.\mathbf{A})\,(\psi_0 + \psi_1 + \psi_2 + \ldots),$$

leading to the series

$$\left(k\frac{\partial}{\partial t} + ii\,\sigma.\nabla + ijm\right)\psi_0 = 0.$$

$$\left(k\frac{\partial}{\partial t} + ii\,\sigma.\nabla + ijm\right)\psi_1 = -e\,(ik\phi - i\,\sigma.\mathbf{A})\,\psi_0$$

$$\left(k\frac{\partial}{\partial t} + ii\,\sigma.\nabla + ijm\right)\psi_2 = -e\,(ik\phi - i\,\sigma.\mathbf{A})\,\psi_1 \ldots .$$

Expanding $(ik\phi - i\,\sigma.\mathbf{A})$ as a Fourier series, and summing over $\mathbf{k}$, we obtain

$$(ik\phi - i\,\sigma.\mathbf{A}) = \Sigma\,(ik\phi(\mathbf{k}) - i\,\sigma.\mathbf{A}(\mathbf{k}))\,e^{i\mathbf{k}.\mathbf{r}},$$

so that

$$\left(k\frac{\partial}{\partial t} + ii\,\sigma.\nabla + ijm\right)\psi_1 = -e\,\Sigma\,(ik\phi(\mathbf{k}) - i\,\sigma.\mathbf{A}(\mathbf{k}))\,e^{i\mathbf{k}.\mathbf{r}}\,\psi_0$$

$$= -e\,\Sigma\,(ik\phi(\mathbf{k}) - i\,\sigma.\mathbf{A}(\mathbf{k}))\,e^{i\mathbf{k}.\mathbf{r}}\,(kE + ii\sigma.\mathbf{p} + ijm)\,e^{-i(Et - \mathbf{p}.\mathbf{r})}$$

$$= -e\,\Sigma\,(ik\phi(\mathbf{k}) - i\,\sigma.\mathbf{A}(\mathbf{k}))\,(kE + ii\sigma.\mathbf{p} + ijm)\,e^{-i(Et - (\mathbf{p}+\mathbf{k}).\mathbf{r})}.$$

Suppose we expand $\psi_1$ as

$$\psi_1 = \Sigma\,v_1(E, \mathbf{p} + \mathbf{k})\,e^{-i(Et - (\mathbf{p}+\mathbf{k}).\mathbf{r})}.$$

Then

$$\Sigma\left(k\frac{\partial}{\partial t} + ii\,\sigma.\nabla + ijm\right)v_1(E, \mathbf{p} + \mathbf{k})\,e^{-i(Et - (\mathbf{p}+\mathbf{k}).\mathbf{r})}$$

$$= -e\,\Sigma\,(ik\phi(\mathbf{k}) - i\,\sigma.\mathbf{A}(\mathbf{k}))\,(kE + ii\sigma.\mathbf{p} + ijm)\,e^{-i(Et - (\mathbf{p}+\mathbf{k}).\mathbf{r})}.$$

$$\Sigma\,(kE + ii\sigma.(\mathbf{p}+\mathbf{k}) + ijm)\,v_1(E, \mathbf{p} + \mathbf{k})\,e^{-i(Et - (\mathbf{p}+\mathbf{k}).\mathbf{r})}$$

$$= -e\,\Sigma\,(ik\phi(\mathbf{k}) - i\,\sigma.\mathbf{A}(\mathbf{k}))\,(kE + ii\sigma.\mathbf{p} + ijm)\,e^{-i(Et - (\mathbf{p}+\mathbf{k}).\mathbf{r})}.$$

and, equating individual terms,

$$(kE + ii\sigma.(\mathbf{p} + \mathbf{k}) + ijm)\ v_1(E, \mathbf{p} + \mathbf{k}) = -e\ (i\,k\phi(\mathbf{k}) - i\,\sigma.\mathbf{A}(\mathbf{k}))\ (kE + ii\sigma.\mathbf{p} + ijm)\ .$$

We can write this in the form

$$v_1(E, \mathbf{p} + \mathbf{k}) = -e\ [kE + ii\,\sigma.(\mathbf{p} + \mathbf{k}) + ijm]^{-1}\ (i\,k\phi(\mathbf{k}) - i\,\sigma.\mathbf{A}(\mathbf{k}))\ (kE + ii\sigma.\mathbf{p} + ijm)\ ,$$

which means that

$$\psi_1 = -e\ \Sigma\ [kE + ii\,\sigma.(\mathbf{p} + \mathbf{k}) + ijm]^{-1}\ (i\,k\phi(\mathbf{k}) - i\,\sigma.\mathbf{A}(\mathbf{k}))$$

$$\times (kE + ii\sigma.\mathbf{p} + ijm)\ e^{-i(Et - (\mathbf{p} + \mathbf{k}).\mathbf{r})}\ .$$

This is the wavefunction for first-order coupling, with an electron (for example) absorbing or emitting a photon of momentum **k**. Applying this to the second term in the perturbation series, we obtain:

$$\left(k\frac{\partial}{\partial t} + i\,i\,\sigma.\nabla + ijm\right)\psi_2 = -e\ (i\,k\phi(\mathbf{k}) - i\,\sigma.\mathbf{A}(\mathbf{k}))\ e^{i\,\mathbf{k}.\mathbf{r}}\ \psi_0\ \times$$

$$\Sigma\ (-e)\ [kE + ii\,\sigma.(\mathbf{p} + \mathbf{k}) + ijm]^{-1}(i\,k\phi(\mathbf{k}) - i\,\sigma.\mathbf{A}(\mathbf{k}))\ (kE + ii\sigma.\mathbf{p} + ijm)\ e^{-i(Et - (\mathbf{p} + \mathbf{k}).\mathbf{r})}\ .$$

We now introduce $\Sigma\ (i\,k\phi(\mathbf{k'}) - i\,\sigma.\mathbf{A}(\mathbf{k'}))\ e^{i\mathbf{k'}.\mathbf{r}}$ for the 4-potential involved with this term, and summing over **k** and **k'**,

$$\psi_2 = \Sigma\ \Sigma\ v_2(E, \mathbf{p} + \mathbf{k} + \mathbf{k'})\ e^{-i(Et - (\mathbf{p} + \mathbf{k} + \mathbf{k'}).\mathbf{r})}\ ,$$

and obtain

$$\Sigma\left(k\frac{\partial}{\partial t} + i\,i\,\sigma.\nabla + ijm\right)v_2(E, \mathbf{p} + \mathbf{k} + \mathbf{k'})\ e^{-i(Et - (\mathbf{p} + \mathbf{k} + \mathbf{k'}).\mathbf{r})} =$$

$$-e\ \Sigma\ \Sigma\ (i\,k\phi(\mathbf{k'}) - i\,\sigma.\mathbf{A}(\mathbf{k'}))\ [kE + ii\,\sigma.(\mathbf{p} + \mathbf{k}) + ijm]^{-1}$$

$$\times (-e)\ (i\,k\phi(\mathbf{k}) - i\,\sigma.\mathbf{A}(\mathbf{k}))\ (kE + ii\sigma.\mathbf{p} + ijm)\ e^{-i(Et - (\mathbf{p} + \mathbf{k} + \mathbf{k'}).\mathbf{r})}\ .$$

Comparing coefficients, we find that

$$(kE + ii\,\sigma.(\mathbf{p} + \mathbf{k} + \mathbf{k'}) + ijm)\ v_2(E, \mathbf{p} + \mathbf{k} + \mathbf{k'}) = -e\ (i\,k\phi(\mathbf{k'}) - i\,\sigma.\mathbf{A}(\mathbf{k'}))$$

$$\times [kE + ii\,\sigma.(\mathbf{p} + \mathbf{k}) + ijm]^{-1}\ (-e)\ (i\,k\phi(\mathbf{k}) - i\,\sigma.\mathbf{A}(\mathbf{k}))\ (kE + ii\sigma.\mathbf{p} + ijm)\ .$$

Hence

$$v_2(E, \mathbf{p} + \mathbf{k} + \mathbf{k'}) = [kE + i i\, \sigma.(\mathbf{p} + \mathbf{k} + \mathbf{k'}) + ijm]^{-1} (-e)\, (i\, k\phi(\mathbf{k'}) - i\, \sigma.\mathbf{A}(\mathbf{k'}))$$

$$\times [kE + i i\, \sigma.(\mathbf{p} + \mathbf{k}) + ijm]^{-1} (-e)\, (i\, k\phi(\mathbf{k}) - i\, \sigma.\mathbf{A}(\mathbf{k}))\, (kE + i i\sigma.\mathbf{p} + ijm)\,,$$

and

$$\psi_2 = \Sigma \Sigma\, [kE + i i\, \sigma.(\mathbf{p} + \mathbf{k} + \mathbf{k'}) + ijm]^{-1} (-e)\, (i\, k\phi(\mathbf{k'}) - i\, \sigma.\mathbf{A}(\mathbf{k'}))$$

$$\times [kE + i i\, \sigma.(\mathbf{p} + \mathbf{k}) + ijm]^{-1} (-e)\, (i\, k\phi(\mathbf{k}) - i\, \sigma.\mathbf{A}(\mathbf{k}))$$

$$\times (kE + i i\sigma.\mathbf{p} + ijm)\, e^{-i(Et - (\mathbf{p} + \mathbf{k} + \mathbf{k'}).\mathbf{r})}\,.$$

This is the wavefunction for second-order coupling, with an electron (for example) absorbing and / or emitting two photons of momenta **k** and **k'**.

## 4 Integral solutions of the Dirac equation

We may suppose that the equation

$$\left(k\frac{\partial}{\partial t} + i i\, \sigma.\nabla + ijm\right)\psi_1 = -e\, (i\, k\phi - i\, \sigma.\mathbf{A})\, \psi_0$$

has an integral solution of the form

$$\psi_1(x) = \int_{-\infty}^{\infty} G_1(x, x')\, \psi_0(x')\, dx\,,$$

where $x$ and $x'$ are 4-vectors, and

$$G_1(x, x') = \frac{1}{(2\pi)^4} \int_{-\infty}^{\infty} d^4p\, [kE + i i\, \sigma.\mathbf{p} + ijm]^{-1} (-e)\, (i\, k\phi(x') - i\, \sigma.\mathbf{A}(x'))\, \exp -ip(x - x')\,,$$

where $p$ is also a 4-vector. Defining the Fourier transform of $(i\, k\phi(x') - i\, \sigma.\mathbf{A}(x'))$ as

$$\frac{1}{(2\pi)^4} \int_{-\infty}^{\infty} d^4p'\, (i\, k\phi(p') - i\, \sigma.\mathbf{A}(p'))\, \exp -ip'(x - x')\,,$$

we obtain

$$G_1(x, x') = \frac{1}{(2\pi)^8} \int\int d^4p\, d^4p'\, [kE + i i\, \sigma.\mathbf{p} + ijm]^{-1}$$

$$\times (-e)\, (i\, k\phi(p') - i\, \sigma.\mathbf{A}(p'))\, \exp -i(p + p')(x - x')\,.$$

The procedure can be extended to $\psi_2(x)$ and

$$G_2(x'', x') = \frac{1}{(2\pi)^8} \int\int d^4p''\, d^4p'''\, [kE'' + i i\, \sigma.\mathbf{p}'' + ijm]^{-1}$$

$$\times (-e)\, (i\, k\phi(p''') - i\, \sigma.\mathbf{A}(p'''))\, \exp -i(p'' + p''')(x'' - x')\,.$$

Reversing the order of *x* and *x'* in the first term, the product of $G_2G_1$ becomes

$$G_2G_1 = \frac{1}{(2\pi)^{16}} \int\int\int\int d^4p\, d^4p'\, d^4p''\, d^4p''' [kE'' + ii\,\sigma.\mathbf{p''} + ijm]^{-1}$$

$$\times (-e)\,(i\,k\phi\,(p''') - i\,\sigma.\mathbf{A}\,(p''')) [kE + ii\,\sigma.\mathbf{p} + ijm]^{-1}$$

$$\times (-e)\,(i\,k\phi\,(p') - i\,\sigma.\mathbf{A}\,(p')) \exp -i(p'' + p''')\,(x'' - x')\, \exp -i(p + p')\,(x' - x).$$

Here, $[kE'' + ii\,\sigma.\mathbf{p''} + ijm]^{-1}\,(-e)\,(i\,k\phi\,(p''') - i\,\sigma.\mathbf{A}\,(p''')) [kE + ii\,\sigma.\mathbf{p} + ijm]^{-1}\,(-e)\,(i\,k\phi\,(p') - i\,\sigma.\mathbf{A}\,(p'))$ has the same form as $v_2(E, \mathbf{p} + \mathbf{k} + \mathbf{k'})$.

## 5 Renormalization

In standard QED, renormalization is not needed at the tree level, where there are no loops, but is needed when loops are required in the calculation. There are essentially three types of graph which remain divergent after the application of gauge invariance and Furry's theorem. These are the electron self-energy graph (with divergence $D = 1$), the electron-photon vertex graph (with $D = 0$), and the photon vacuum polarization graph (with $D = 2$, which can be reduced to $D = 0$, by gauge invariance). Let us take the electron self-energy graph, for a free electron not interacting with any external field. In effect, this is a representation of an electron emitting a virtual photon and then reabsorbing it. The perturbation expansion for a first-order coupling of this kind (for either emission or absorption) produces a wavefunction of the form

$$\Psi_1 = -e\,\Sigma\,[kE + ii\,\sigma.(\mathbf{p} + \mathbf{k}) + ijm]^{-1}\,(ik\phi - i\,\sigma.\mathbf{A})\,(kE + ii\sigma.\mathbf{p} + ijm)\,e^{-i(Et - (\mathbf{p} + \mathbf{k}).\mathbf{r})},$$

where **k** is the extra momentum acquired by the electron from the photon. If we observe the process in the rest frame of the electron and eliminate any external source of potential, then $\mathbf{k} = 0$, and $(ik\phi - i\,\sigma.\mathbf{A})$ reduces to the static value, $ik\phi$. In this case, $\Psi_1$ becomes

$$\Psi_1 = -e\,\Sigma\,[kE + ii\,\sigma.\mathbf{p} + ijm]^{-1}\,ik\phi\,(kE + ii\sigma.\mathbf{p} + ijm)\,e^{-i(Et - (\mathbf{p} + \mathbf{k}).\mathbf{r})}.$$

Writing this as

$$\Psi_1 = -e\,\Sigma\,[kE + ii\,\sigma.\mathbf{p} + ijm]^{-1}\,[-kE + ii\,\sigma.\mathbf{p} + ijm]^{-1}\,(-kE + ii\,\sigma.\mathbf{p} + ijm)\,ik\phi$$

$$\times (kE + ii\sigma.\mathbf{p} + ijm)\,e^{-i(Et - (\mathbf{p} + \mathbf{k}).\mathbf{r})},$$

we obtain $\Psi_1 = 0$, for any fixed value of $\phi$. In other words, a *non-interacting* electron requires no renormalization as a result of its self-energy. We can extend this idea to the photon vacuum polarization, which creates and then annihilates an electron-positron pair. These processes are equivalent, in effect to producing a diagram in which an electron and positron travel in mutual opposite directions from or towards a vertex with an incoming or outgoing photon, which, in the absence of any external potential, is, in principle, the same as an electron emitting and absorbing a photon and being undeviated by any momentum change ($\mathbf{k} = 0$), as in the self-energy diagram. The calculation of $\Psi_1 = 0$ for the self-energy should, therefore, also apply in this case. The electron-photon vertex graph diverges only beyond the tree level, when a photon loop links the electron on each side of a vertex formed with an incoming photon. In the context of $\mathbf{k} = 0$, this reduces to another equivalent of electron self-energy or vacuum polarization.

It appears, from this calculation that $[kE + i\mathbf{i}\,\sigma.\mathbf{p} + i\mathbf{j}m]^{-1}$ is equivalent to the vacuum 'image' of $(kE + i\mathbf{i}\,\sigma.\mathbf{p} + i\mathbf{j}m)$, that is, $(-kE + i\mathbf{i}\,\sigma.\mathbf{p} + i\mathbf{j}m)$, and this seems to be its physical meaning. In the nilpotent formulation, a fermion with quaternion state vector $(kE + i\mathbf{i}\,\sigma.\mathbf{p} + i\mathbf{j}m)$ generates a vacuum image of the form $(-kE + i\mathbf{i}\,\sigma.\mathbf{p} + i\mathbf{j}m)$ by acting on the vacuum operator $k(-kE + i\mathbf{i}\,\sigma.\mathbf{p} + i\mathbf{j}m)$. However, $(kE + i\mathbf{i}\,\sigma.\mathbf{p} + i\mathbf{j}m)$ $k$ $(kE + i\mathbf{i}\,\sigma.\mathbf{p} + i\mathbf{j}m)$ is, in principle identical to $(kE + i\mathbf{i}\,\sigma.\mathbf{p} + i\mathbf{j}m)$, as is $(kE + i\mathbf{i}\,\sigma.\mathbf{p} + i\mathbf{j}m)$ $k$ $(kE + i\mathbf{i}\,\sigma.\mathbf{p} + i\mathbf{j}m)$ $k$ $(kE + i\mathbf{i}\,\sigma.\mathbf{p} + i\mathbf{j}m)$ $k$ $(kE + i\mathbf{i}\,\sigma.\mathbf{p} + i\mathbf{j}m)$ …, in an infinite series of actions upon the vacuum; and this is mathematically and physically equivalent to $(kE + i\mathbf{i}\,\sigma.\mathbf{p} + i\mathbf{j}m)$ $(-kE + i\mathbf{i}\,\sigma.\mathbf{p} + i\mathbf{j}m)$ $(kE + i\mathbf{i}\,\sigma.\mathbf{p} + i\mathbf{j}m)$ $(-kE + i\mathbf{i}\,\sigma.\mathbf{p} + i\mathbf{j}m)$ …, which generates an infinite series of alternate fermion and boson states. We can, therefore, conceive of this in terms of a fermion generating an alternate series of boson and fermion loops, which sum up to return to the pure fermion state itself. It is effectively a form of supersymmetry in which the supersymmetric partners are not new particles but merely vacuum images or couplings of the original state.

In addition to this, as we have shown in earlier work, the nilpotent representation of the Dirac equation is already second quantized because the operator and state vector are necessarily identical, and each is quantized. This means that the quantum field integrals reduce in principle to the form of a single quaternion state vector.[1] This representation says nothing directly about the energy value associated with the coupling of a charge to an electromagnetic field, but it does define the fact that such a single representation must be possible, and, that the infinite number of possible fluctuations must be constrained in such a way that they sum up to a single finite value for the coupling in any given state. From this perspective, the existence of a method for eliminating apparent infinities is no more surprising than the fact that complicated physical systems act in such a way that they conserve energy or that the complicated mess of quarks and gluons inside a proton somehow 'conspires' to deliver a total spin value of ½ to the composite particle. It is also not surprising that there is a maximum value of energy involved which ensures that the summation is finite. The existence of the quaternion state vector as a single representation of the entire quantum field is a statement that a finite summation exists. The calculation at the beginning of this section is an illustration of the fact. There is even a mechanism for ensuring that it happens in the interactions of quaternion state vectors with the vacuum.

When we wish to determine the relative strength of the coupling involved in an electron interacting with an external field, for example in the process of electron-muon scattering or the coupling of the electron to a vector potential to determine its magnetic moment, we *then* need to sum a perturbative series of real interacting terms. The process, however, will not be divergent, and it will naturally cut off at the value of energy described as the Planck mass ($M_P$), the energy at which the effects of what is usually referred to as quantum gravity take place. This is, we believe, because gravity is the carrier of the wavefunction correlations involved in nonlocality, and the Planck mass is the quantum of the inertial interactions which result from the effect of gravity on the time-delayed nature of the nongravitational interactions, which, in turn, produces the inertial mass associated with 'charged' particles. It is, of course, the fact that such particles have inertial mass which creates the time delay in the transmission of energy; and the inertial mass may also be seen as the result of a coupling to the Higgs field, which produces the filled vacuum which allows the instantaneous transmission of gravity and wavefunction correlations. Cutting off at the Planck mass (($\hbar c / G)^{1/2}$), which is *c*-dependent is the same thing, in effect, as saying that the real interactions (and inertial reactions) are time-delayed or that the fermions involved have inertial mass, and it is significant that, in QED, the process actually determines the mass-value of the electron. The process is, in effect, creating a *c*-related 'event horizon' such as we would expect for an inertial process as described.

The significant result from current QED calculations is that the first-order correction term for the electric fine structure constant, for purely leptonic interactions, at energy of interaction $\mu$, is $(1 / 3\pi) \ln (M_P^2 / \mu^2)$, which has the same form as the term which we have derived for the quark theory, using the idea of lepton-type quarks, and also the correction term which is needed to achieve Grand Unification at the Planck Mass. The reasons for the Planck mass being the energy for a $U(5)$ unification, involving gravity, have been discussed in earlier papers, together with relevant predictions.[3,5]

## 6 Green's function solution

To apply the method of Green's functions, we solve for a unit source and then sum over the whole distribution. Beginning with

$$\left(k \frac{\partial}{\partial t} + i\, i\, \sigma.\nabla + ijm\right)\psi = -e\,(i\,k\phi - i\,\sigma.\mathbf{A})\,\psi\;,$$

or

$$\left(k \frac{\partial}{\partial t} + i\, i\, \sigma.\nabla + ijm\right)(kE + ii\sigma.\mathbf{p} + ijm)e^{ip.x} = -e(i\,k\phi - i\,\sigma.\mathbf{A})\,(kE + ii\sigma.\mathbf{p} + ijm)e^{ip.x}\;,$$

where *p* and *x* are 4-vectors, we solve for a unit source using

$$\left(k \frac{\partial}{\partial t} + i\, i\, \sigma.\nabla + ijm\right)G_F = \delta^{(4)}(x - x')\;,$$

where $G_F$ is a wave produced at $x$ by a unit source at $x'$. Translational invariance shows that $G_F$ is a function of $(x - x')$. Then the solution for the whole distribution becomes

$$\psi(x) = -e \int d^4x' \; G_F(x - x') \; (i\, k\phi - i\, \sigma.\mathbf{A}) \; \psi(x')$$

or

$$(kE + ii\sigma.\mathbf{p} + ijm)e^{ip.x} = e \int d^4x' \; G_F(x - x') \; (i\, k\phi - i\, \sigma.\mathbf{A}) \; (kE + ii\sigma.\mathbf{p} + ijm)e^{ip.x'}.$$

Taking the Fourier transform into momentum space, we obtain

$$G_F(x - x') = \frac{1}{(2\pi)^4} \int d^4p \; S_F(p) \; e^{-ip.(x - x')}.$$

Then, using the Fourier representation of the $\delta$-function, we find

$$\frac{1}{(2\pi)^4} \int d^4p \left( k \frac{\partial}{\partial t} + i\,i\,\sigma.\nabla + ijm \right) S_F(p) \; e^{-ip.(x - x')} = \frac{1}{(2\pi)^4} \int d^4p \; e^{-ip.(x - x')},$$

from which

$$(kE + ii\sigma.\mathbf{p} + ijm) \, S_F(p) = 1$$

and

$$S_F(p) = \frac{1}{(kE + ii\sigma.\mathbf{p} + ijm)},$$

which is the electron propagator in the Feynman formalism. Conventional theory assumes that

$$S_F(p) = \frac{1}{\not{p} - m} = \frac{\not{p} + m}{p^2 - m^2},$$

and that there is a singularity or 'pole' ($p_0$) where $p^2 - m^2 = 0$, the 'pole' being the origin of positron states. Effectively, on either side of the pole we have positive energy states moving forwards in time, and negative energy states moving backwards in time, the terms $(\not{p} + m)$ and $(-\not{p} + m)$ being used to project out, respectively, the positive and negative energy states. We then add the infinitesimal term $i\varepsilon$ to $p^2 - m^2$, so that $iS_F(p)$ becomes

$$iS_F(p) = \frac{i(\not{p} + m)}{p^2 - m^2 + i\varepsilon},$$

$$= \frac{(\not{p} + m)}{2p_0} \left( \frac{1}{p_0 - \sqrt{p^2 + m^2} + i\varepsilon} + \frac{1}{p_0 + \sqrt{p^2 + m^2} - i\varepsilon} \right),$$

and take a contour integral over the complex variable to give the solution

$$S_F(x - x') = \int d^3p \, \frac{1}{(2\pi)^3} \frac{m}{2E} \, [-i\theta(t - t') \sum_{r=1}^{2} \Psi(x) \overline{\Psi}(x')$$

$$+ i\theta(t' - t) \sum_{r=3}^{4} \Psi(x) \overline{\Psi}(x')],$$

with summations over the up and down spin states.

In the nilpotent formalism, however, the $i\varepsilon$ term is unnecessary and there is no infrared divergence at the pole, because the denominator of the propagator term is a positive nonzero scalar. We write

$$S_F(p) = \frac{1}{(kE + ii\sigma.\mathbf{p} + ijm)} ,$$

and are free to choose

$$\frac{1}{(kE + ii\sigma.\mathbf{p} + ijm)} = \frac{(-kE + ii\sigma.\mathbf{p} + ijm)}{(kE + ii\sigma.\mathbf{p} + ijm)(-kE + ii\sigma.\mathbf{p} + ijm)}$$

$$= \frac{(-kE + ii\sigma.\mathbf{p} + ijm)}{(E^2 + p^2 + m^2)} ,$$

which is finite at all values. We can also, for example, write

$$\frac{1}{(kE + ii\sigma.\mathbf{p} + ijm)} = \frac{-kE + ii\sigma.\mathbf{p} + ijm}{-2kE} \left( \frac{1}{(kE + ii\sigma.\mathbf{p} + ijm)} - \frac{1}{(-kE + ii\sigma.\mathbf{p} + ijm)} \right)$$

and other, similar, forms. Our integral is now simply

$$S_F(x - x') = \int d^3p \, \frac{1}{(2\pi)^3} \frac{m}{2E} \, \theta(t - t') \, \Psi(x) \, \overline{\Psi}(x') ,$$

in which

$$\Psi(x) = ((kE + ii\,\mathbf{p} + ij\,m) \ldots) \exp(ipx) ,$$

where $((kE + ii\,\mathbf{p} + ij\,m) \ldots)$ is the bra matrix with the terms:

$$(kE + ii\,\mathbf{p} + ij\,m); (kE - ii\,\mathbf{p} + ij\,m); (-kE + ii\,\mathbf{p} + ij\,m); (-kE - ii\,\mathbf{p} + ij\,m) ,$$

and the adjoint term becomes

$$\overline{\Psi}(x') = ((kE - ii\,\mathbf{p} - ij\,m) \ldots)(ik) \exp(-ipx') ,$$

with $((kE - ii\,\mathbf{p} - ij\,m) \ldots)(ik)$ a ket. The reason for this success is apparent. The nilpotent formulation is automatically second quantized and the negative energy states appear as components of the nilpotent wavefunction on the same basis as the positive energy states. No averaging over spin states or 'interpreting' $-E$ as a reversed time state is necessary; the 'reversed time' state occurs with the $t$ in the operator $\partial / \partial t$, and there is no need to separate out the states on opposite sides of the pole.

We can use the electron propagator to define the photon propagator. Conventionally, we derive the photon propagator, $\Delta_F(x - x')$, directly from the Klein-Gordon equation, while recognizing that its mathematical form depends on the choice of gauge. In the presence of a source field, represented by current $j(x)$, we can write the Klein-Gordon equation in the form

$$\left(\frac{\partial^2}{\partial t^2} - \nabla^2 - m^2\right)\phi(x) = j(x),$$

and, with $\Delta_F(x - x')$ as Green's function, we have

$$\left(\frac{\partial^2}{\partial t^2} - \nabla^2 - m^2\right)\Delta_F(x - x') = -\delta^{(4)}(x - x'),$$

for which the solution is

$$\phi(x) = \phi_0(x) - \int d^4x' \, \Delta_F(x - x') \, j(x'),$$

where $\phi_0(x)$ is a solution of the equation in the absence of sources:

$$\left(\frac{\partial^2}{\partial t^2} - \nabla^2 - m^2\right)\phi(x) = 0.$$

Using the Fourier transform

$$\Delta_F(x - x') = \frac{1}{(2\pi)^4} \int d^4p \, \exp{-ip(x - x')} \, \Delta_F(p),$$

and applying the operator $\left(\frac{\partial^2}{\partial t^2} - \nabla^2 - m^2\right)$ to both sides, we obtain:

$$\Delta_F(x - x') = \frac{\not{p} + m}{p^2 - m^2}.$$

Clearly, the relationship of the electron and photon propagators is of the form

$$S_F(x - x') = (i \, \gamma^\mu \partial_\mu + m) \, \Delta_F(x - x'),$$

or, in our notation,

$$S_F(x - x') = ((\mathbf{k}E + i\mathbf{i} \, \mathbf{p} + i\mathbf{j} \, m) \ldots) \, \Delta_F(x - x').$$

This is exactly what we would expect in transferring from boson (Klein-Gordon field) to fermion (Dirac field), using our single vector operator. Now, using

$$iS_F(p) = \frac{1}{2p_0}\left(\frac{1}{p_0 - \sqrt{p^2 + m^2} + i\varepsilon} + \frac{1}{p_0 + \sqrt{p^2 + m^2} - i\varepsilon}\right),$$

which is the same as the conventional electron propagator up to a factor $(\not{p} + m)$, we can perform virtually the same contour integral as in the case of the electron to produce

$$i\Delta_F(x - x') = \int d^3p \, \frac{1}{(2\pi)^3} \frac{1}{2\omega} \, \theta(t - t') \, \phi(x)\phi^*(x'),$$

where $\omega$ takes the place of $E / m$. This time, of course, $\phi(x)$ and $\phi(x')$ are scalar wavefunctions. In our notation, they are each scalar products of the 4-component bra term (($kE + i\boldsymbol{i}\ \mathbf{p} + i\boldsymbol{j}\ m$) …) and the 4-component ket term ((–$kE + i\boldsymbol{i}\ \mathbf{p} + i\boldsymbol{j}\ m$) …), multiplied respectively by exponentials exp ($ipx$) and exp ($ipx'$), expressed in terms of the 4-vectors $p$, $x$ and $x'$. In the nilpotent formulation, $\phi(x)\phi^*(x')$ reduces to a product of a scalar term (which can be removed by normalization) and exp $ip(x – x')$.

Now, the product $\Psi(x)\ \overline{\Psi}(x')$, for the electron propagator, can be interpreted as the product of (($kE + i\boldsymbol{i}\ \mathbf{p} + i\boldsymbol{j}\ m$) …) exp ($ipx$) and the 'vacuum operator', $i\boldsymbol{k}$ (($kE + i\boldsymbol{i}\ \mathbf{p} + i\boldsymbol{j}\ m$) …) exp (–$ipx'$), which reduces to the product of a scalar (again removed by normalization), (($kE + i\boldsymbol{i}\ \mathbf{p} + i\boldsymbol{j}\ m$) …) and exp $ip(x – x')$. The integrals, like the propagators, are related by the factor (($kE + i\boldsymbol{i}\ \mathbf{p} + i\boldsymbol{j}\ m$) …), which suggests that we should *define* the photon propagator, in terms of the electron propagator, for which, in the nilpotent formulation, we can do a completely non-divergent integral without additional infinitesimal terms, rather than as a result of the Klein-Gordon equation. In principle, this must reflect the fact that photon processes cannot ultimately be described independently of processes involving paired fermions or fermions and antifermions (one of which may be manifested indirectly in terms of a potential, rather than directly as a charge). We can then write

$$(kE + i\boldsymbol{i}\ \mathbf{p} + i\boldsymbol{j}\ m)\ \Delta_F\ (x – x') = S_F\ (x – x') = (–kE + i\boldsymbol{i}\ \mathbf{p} + i\boldsymbol{j}\ m)\ ,$$

which means that (after normalization) $S_F\ (x – x')$ takes the expected form of $q^{-2} = 1 / (E^2 – p^2)$, with $m^2 = 0$. The apparently divergent nature of the photon propagator integral, in conventional theory, is simply a reflection of the fact that photons, unlike electrons, have no independent existence and are not conserved objects. However, if we define the photon propagator in terms of the nilpotent fermion propagator integral, we will be using the automatic cancellation process which the nilpotent representation of the fermion introduces.

**7 The propagator method in lowest order**

The Dirac equation in the presence of external Coulomb potentials,

$$\left(k\frac{\partial}{\partial t} + i\boldsymbol{i}\ \sigma.\nabla + ijm\right)\psi = -e\ (i\ \boldsymbol{k}\phi - \boldsymbol{i}\ \sigma.\mathbf{A})\ \psi$$

may be solved by introducing the Dirac propagator, so that

$$\left(k\frac{\partial}{\partial t} + i\boldsymbol{i}\ \sigma.\nabla + ijm\right)S_F\ (x – x') = -e\ (i\ \boldsymbol{k}\phi - \boldsymbol{i}\ \sigma.\mathbf{A})\ \delta^{(4)}\ (x – x')\ .$$

Then

$$\psi(x) = \psi_0(x) + \int d^4x'\ S_F\ (x – x')\ (-e)\ (i\ \boldsymbol{k}\phi - \boldsymbol{i}\ \sigma.\mathbf{A})\ \psi\ ,$$

where $\psi_0(x)$ is the solution of

$$\left(k\frac{\partial}{\partial t} + i\,i\,\sigma.\nabla + ijm\right)\psi_0 = 0 \ .$$

The Dirac equation with potentials may also be written in terms of the Hamiltonian,

$$H = i\,k\,e\phi + i\,i\,\sigma.(\nabla + i\,e\mathbf{A}) + ijm = H_0 + H_1 \ ,$$

where

$$H_0 = i\,i\,\sigma.\nabla + ijm \ ,$$

and $H_1$ is the interaction term,

$$H_1 = i\,k\,e\phi - i\,e\,\sigma.\mathbf{A}) \ .$$

Then

$$\left(k\frac{\partial}{\partial t} + i\,H\right)\psi = 0$$

and

$$\left(k\frac{\partial}{\partial t} + i\,H_0 + i\,H_1\right)G(x - x') = \delta^{(4)}(x - x') \ .$$

Solving the Green's function, $G(x - x')$, for the general case involving interactions, allows us to use Huygens' principle to find the time evolution of the wave, by summing all the secondary wavelets produced by infinitesimal point sources on the wavefront. In mathematical terms (changing over the variables from 4-vector notation),

$$\psi(\mathbf{x}, t) = \int d^3x'\, G_F(\mathbf{x} - \mathbf{x}', t - t')\, \psi(\mathbf{x}', t') \ .$$

Using the symbolic identity

$$G = G_0 + G_0 H_1 G_0 + G_0 H_1 G_0 H_1 G_0 + \ldots ,$$

and a power expansion, and applying Huygens' principle, the time evolution of the wavefunction becomes described by

$$\psi(\mathbf{x}, t) = \psi_0(\mathbf{x}, t) + \int d^4x_1 G_0(\mathbf{x} - \mathbf{x_1}, t - t_1)\, H_1(\mathbf{x_1}, t_1)\, \psi_0(\mathbf{x_1}, t_1)$$

$$+ \int d^4x_1\, d^4x_2\, G_0(\mathbf{x} - \mathbf{x_1}, t - t_1)\, H_1(\mathbf{x_1}, t_1)\, G_0(\mathbf{x_1} - \mathbf{x_2}, t_1 - t_2)\, \psi_0(\mathbf{x_2}, t_2) + \ldots .$$

To calculate the transition probability, in lowest order, that a free plane wave in initial state $i$ emerges in final state $f$ after scattering off a potential, we multiply the equation

$$\psi(\mathbf{x}, t) = \phi_i(\mathbf{x}, t) + \int d^4x'\, G_0(\mathbf{x} - \mathbf{x}', t - t')\, H_1(\mathbf{x}', t')\, \phi_i(\mathbf{x}', t')$$

on the left by $\phi_j^*$ and integrate, obtaining the scattering matrix,

$$S_{fi} = \delta_{fi} + i \int d^4x\, \phi_f^*(x')\, H_1(x')\, \phi_i(x') + \ldots$$

Considering, now, an electron interacting with external Coulomb potentials, we introduce

$$\psi(x) = \psi_0(x) + \int d^4x'\, S_F(x-x')\, (-e)\, (i\,k\phi - i\,\sigma.\mathbf{A})\, \psi,$$

and obtain

$$\psi(x) = \psi_i(x) + (-e)\int d^4x'\, \theta(t-t') \int d^3p\, \frac{1}{(2\pi)^3}\, \frac{m}{2E}\, \psi(x)\, \bar{\psi}(x')(i\,k\phi(x') - i\,\sigma.\mathbf{A}(x'))\, \psi(x').$$

Multiplying both sides of the equation from the left by $\bar{\psi}_f$ and integrating over all space and time, we obtain the scattering matrix, to lowest order:

$$S_{fi} = \delta_{fi} - ie \int \bar{\psi}_f\, \gamma_\mu A_\mu\, \psi_i\, d^4x.$$

**8 Electron scattering**

From the previous section, we find that the amplitude for the scattering of an electron from initial state $\psi_i$ to final state $\psi_f$ is given by

$$T_{fi} = ie \int \bar{\psi}_f\, \gamma_\mu A_\mu\, \psi_i\, d^4x$$

$$= ie \int i\mathbf{k}\, (kE_f + ii\sigma.\mathbf{p}_f + ijm_f)\, \exp(i\,(E_f t - \mathbf{p}_f.\mathbf{r}))$$

$$\times (ik\phi - i\,\sigma.\mathbf{A})\, (kE_i + ii\sigma.\mathbf{p}_i + ijm_i)\, \exp(i\,(E_i t - \mathbf{p}_i.\mathbf{r}))\, d^3r\, dt$$

$$= ie \int i\mathbf{k}\, (kE_f + ii\sigma.\mathbf{p}_f + ijm_f)\, (ik\phi - i\,\sigma.\mathbf{A})$$

$$\times (kE_i + ii\sigma.\mathbf{p}_i + ijm_i)\, \exp(-i\,((E_i - E_f)t - (\mathbf{p}_i - \mathbf{p}_f).\mathbf{r}))\, d^3r\, dt.$$

It is convenient to write this in terms of the electromagnetic transition current between the initial and final states,

$$j^\mu_{fi} = -e\, \bar{\psi}_f\, \gamma_\mu\, \psi_i,$$

but this is only possible if we redefine the scalar product $\gamma_\mu A^\mu$ as a product of diagonal matrix ($\gamma_\mu$) and column vector ($A^\mu$), to separate out the terms $i\mathbf{k}$ and $-i\sigma$, within $\gamma_\mu$, which have quite different actions on $(kE_i + ii\sigma.\mathbf{p}_i + ijm_i)$. In effect, $\mathbf{k}$ reverses the sign of $E$ (time-reversal) and $i$ reverses the sign of $\mathbf{p}$ (parity), so transforming $\gamma_\mu$ into a matrix is like applying the metric tensor, $g^{\mu\nu}$. Thus, the row vector $\gamma_\mu$ now becomes a diagonal matrix of the form:

$$\begin{pmatrix} ik & 0 & 0 & 0 \\ 0 & -i\sigma_1 & 0 & 0 \\ 0 & 0 & -i\sigma_2 & 0 \\ 0 & 0 & 0 & -i\sigma_3 \end{pmatrix}$$

It is convenient, also, to write $\exp(-i((E_i - E_f)t - (\mathbf{p}_i - \mathbf{p}_f)\cdot\mathbf{r})) = \exp(-i\,q.x)$, in terms of a 4-vector difference $x$ and a 4-momentum difference $q$, of which the vector part is $\mathbf{q} = \mathbf{p}_i - \mathbf{p}_f$. So, the electron current becomes

$$j^\mu_{fi} = -e\,\bar{\psi}_f\,\gamma_\mu\,\psi_i = -e\,ik\,(kE_f + ii\sigma.\mathbf{p}_f + ijm_f)\,\gamma_\mu\,(kE_i + ii\sigma.\mathbf{p}_i + ijm_i)\,e^{-iq.x}\ .$$

Defining $A^\mu(x)$ as the 4-vector potential associated with the static charge, and $A^\mu(q)$ as its Fourier transform, with vector component $A^\mu(\mathbf{q})$,

$$A^\mu(q) = \int x\,e^{-iq.x}\,A^\mu(x)\,d^4x\ ,$$

the amplitude for the process becomes

$$T_{fi} = ie \int \bar{\psi}_f\,\gamma_\mu\,\psi_i\,A^\mu(x)\,d^4x = -i\int j^\mu_{fi}\,A^\mu(x)\,d^4x\ ,$$

$$= -e\,k\,(kE_f + ii\sigma.\mathbf{p}_f + ijm_f)\,\gamma_\mu\,(kE_i + ii\sigma.\mathbf{p}_i + ijm_i)\,A^\mu(q)\ .$$

For a static source, with time-independent $A^\mu(x)$,

$$A^\mu(q) = \int \exp(-i(E_i - E_f)t)\,dt \int e^{i\mathbf{q}.\mathbf{x}}\,A^\mu(\mathbf{x})\,d^3x\ .$$

$$= 2\pi\,\delta(E_f - E_i)\,A^\mu(\mathbf{q})\ .$$

Applying Maxwell's equations for time-independent $A^\mu(x)$, we have

$$\nabla^2 A^\mu(\mathbf{x}) = -j^\mu(\mathbf{x})\ ,$$

and so

$$\int e^{i\mathbf{q}.\mathbf{x}}\,A^\mu(\mathbf{x})\,d^3x = \int (\nabla^2 A^\mu(\mathbf{x}))\,e^{i\mathbf{q}.\mathbf{x}}\,d^3x$$

$$= -|\mathbf{q}|^2\,A^\mu(\mathbf{q}) = -j^\mu(\mathbf{q})\ .$$

Hence

$$A^\mu(\mathbf{q}) = \frac{1}{|\mathbf{q}|^2}j^\mu(\mathbf{q})\ ,$$

and

$$T_{fi} = -2i\pi\,\delta(E_f - E_i)\,e\,k\,(kE_f + ii\sigma.\mathbf{p}_f + ijm_f)\,\gamma_\mu\,(kE_i + ii\sigma.\mathbf{p}_i + ijm_i)\,\frac{1}{|\mathbf{q}|^2}j^\mu(\mathbf{q})\ .$$

Removing the $\delta$-function, we obtain the covariant amplitude:

$$-i\,M = -e\,k\,(kE_f + ii\sigma.\mathbf{p}_f + ijm_f)\,\gamma_\mu\,(kE_i + ii\sigma.\mathbf{p}_i + ijm_i)\,\frac{1}{|\mathbf{q}|^2}\,j^\mu(\mathbf{q})\;.$$

Applying energy conservation, $E_f = E_i$, and $q^2 = -|\mathbf{q}|^2$, which means that the invariant amplitude may be written:

$$-i\,M = -e\,k\,(kE_f + ii\sigma.\mathbf{p}_f + ijm_f)\,\gamma^\mu\,(kE_i + ii\sigma.\mathbf{p}_i + ijm_i)\,\frac{-ig_{\mu\nu}}{q^2}\,(-ij^\nu(\mathbf{q}))\;.$$

We can, of course, use this equation, for example, with $\gamma_\mu = \gamma_0$ and $g_{\mu\nu} = 1$, and $iZe$ replacing $ij^\nu(\mathbf{q})$, to find a value for the angular distribution ($|M|^2$) in the case of Rutherford scattering off a static nuclear charge $Ze$; and we can also assume that more complicated calculations can be done using the Feynman rules which codify the perturbative method. However, our main concern is with developing an approach to renormalization. Let us suppose, therefore, that the exchanged photon fluctuates into an electron-positron pair. Application of the Feynman rules requires a factor $(-1)^n$ for a diagram containing $n$ fermion loops (in this case, $n = 1$). It also requires an integral over $d^4p\,/\,(2\pi)^4$ to sum over all possible values of the unobservable $p$, in the part of the invariant amplitude referring to the loop. The invariant amplitude is, therefore, now:

$$-i\,M = -(-)^1\,e\,k\,(kE_f + ii\sigma.\mathbf{p}_f + ijm_f)\,\gamma^\mu\,(kE_i + ii\sigma.\mathbf{p}_i + ijm_i)\,\frac{-ig_{\mu\mu'}}{q^2}$$

$$\times \frac{1}{(2\pi)^4}\,\int (ie\gamma^{\mu'})_{\alpha\beta}\,\frac{i(kE + ii\sigma.\mathbf{p} + ijm)_{\beta\lambda}}{(E^2 - p^2 - m^2)}\,((ie\gamma^{\nu'})_{\lambda\tau}\,\frac{i(ii\sigma.(\mathbf{p} - \mathbf{q}) + ijm)_{\beta\lambda}}{((p-q)^2 - m^2)}\,d^4p$$

$$\times \frac{-ig_{\nu'\nu}}{q^2}\,(-ij^\nu(\mathbf{q}))\;.$$

In principle, the addition of this term for the photon loop to the lowest-order invariant amplitude for electron scattering can be achieved by the addition of a modifying term to the lowest-order propagator, so that

$$\frac{-ig_{\mu\nu}}{q^2} \to \frac{-ig_{\mu\mu'}}{q^2} + \frac{-ig_{\nu'\nu}}{q^2}\,I^{\mu'\nu'}\,\frac{-ig_{\mu\nu}}{q^2}$$

$$\to \frac{-ig_{\mu\nu}}{q^2} + \frac{-i}{q^2}\,I_{\mu\nu}\,\frac{-i}{q^2}$$

with

$$I_{\mu\nu}(q^2) = (-)^1\,\frac{1}{(2\pi)^4}\,\int \mathrm{Tr}\left((ie\gamma_\mu)\frac{i(kE + ii\sigma.\mathbf{p} + ijm)}{(E^2 - p^2 - m^2)}\,(ie\gamma_\nu)\frac{i(ii\sigma.(\mathbf{p}-\mathbf{q}) + ijm)}{((p-q)^2 - m^2)}\right)d^4p\;.$$

Bjorken and Drell[6] and others show that, with the omission of terms which disappear when the propagator is coupled to external charges or currents, $I_{\mu\nu}$ can be written as

$$I_{\mu\nu} = -g_{\mu\nu}\,q^2\,I(q^2)$$

with

$$I(q^2) = \frac{\alpha}{3\pi} \int_{m^2}^{\infty} \frac{dp}{p^2} - \frac{2\alpha}{\pi} \int_0^1 z(1-z^2) \ln\left(1 - \frac{q^2 z(1-z^2)}{m^2}\right) dz,$$

where $m$ is the electron mass. In principle, the first term of this integral is divergent, but with $M_P$ representing a natural cut-off value for the mass, $I(q^2)$ becomes the convergent:

$$I(q^2) = \frac{\alpha}{3\pi} \int_{m^2}^{M_P^2} \frac{dp}{p^2} - \frac{2\alpha}{\pi} \int_0^1 z(1-z^2) \ln\left(1 - \frac{q^2 z(1-z^2)}{m^2}\right) d .$$

For small values of $(-q^2)$, we have

$$I(q^2) \approx \frac{\alpha}{3\pi} \ln \frac{M_P^2}{m^2} + \frac{\alpha}{15\pi} \frac{q^2}{m^2},$$

while large values of $(-q^2)$ lead to

$$I(q^2) \approx \frac{\alpha}{3\pi} \ln \frac{M_P^2}{m^2} - \frac{\alpha}{3\pi} \frac{-q^2}{m^2} = \frac{\alpha}{3\pi} \ln \frac{M_P^2}{-q^2} .$$

The first term in each case may be considered as a modification to the fine structure constant or the electric charge value as produced by the lowest-order Feynman diagram. The modified (or measured) charge ($e'$) then becomes:

$$e' = e \left(1 - \frac{e^2}{3\pi} \ln \frac{M_P^2}{m^2}\right)^{1/2} .$$

In principle, of course, 'charge' really represents an existence condition rather than a value. The 'value' of charge is a measure of the coupling to the energy of the field. Charge, as a fundamentally imaginary quantity, has no value. Its 'value' can only be determined by its coupling to another charge, and then is only determined in terms of the energy of the coupling. In this respect, charge is unlike mass, which, as a real quantity, has a value independent of its coupling with other masses. We can only know that a charge exists by its interaction with other charges, but we can know of the existence of a mass through its inertia, irrespective of the existence of other masses. The scaling of the interaction value associated with an electromagnetic charge at different energies is, therefore, an expected process, which is related to a similar scaling of the electron mass (calculated from the self-energy diagram), and stems from the same vacuum origin by which this is created. In the absence of an ultraviolet divergence in the integral, it produces the expected result, demanded by the nilpotent algebra, of a finite (but not independently determined) value of charge at all energies.

It is significant that the alternative (BPHZ) method of renormalization, via the adding of counterterms to the action, results in the same scaling of mass and charge as the direct multiplicative method described above. We can consider this as equivalent to the automatic addition of inherent supersymmetric partners in the nilpotent algebra, which has the advantage of being a required property at each successive order. A similar process of cancelling the contributions from electrons and those from electrons plus very soft (undetectable) photons has been used to remove the infrared divergences in bremsstrahlung and pair production. Such procedures can be related to our analyses of divergent diagrams in section 5, and our discussion of the creation of alternate fermionic and bosonic vacuum 'images' in section 2 (where the additional photon energy is actually zero, and so necessarily undetectable).

**9 Strong and weak analogues**

Both strong and weak interactions have relationships between fermion and boson propagators which follow the same overall pattern as that in QED. The quark propagator in the strong interaction is simply that for any fermion, and the fermion propagator, $S_F(p)$ or $iS_F(p)$, applies to the weak, as well as to the electromagnetic interaction. The conventional

$$iS_F(p) = \frac{i}{\not{p} - m} = \frac{i(\not{p} + m)}{p^2 - m^2}$$

becomes

$$iS_F(p) = \frac{i}{(kE + ii\sigma.\mathbf{p} + ijm)} = \frac{i(-kE + ii\sigma.\mathbf{p} + ijm)}{(E^2 + p^2 + m^2)}$$

in the nilpotent formulation, and its general applicability means that we can eliminate the infrared divergence in the fermion propagators for all of the interactions.

It is interesting that the 'renormalizability' of the combined electroweak interaction is related to the very mechanism which gives masses to the fermions and gauge bosons. For obvious reasons, a quantum field integral taken over all values of $p$ will only be finite, as the nilpotent algebra demands, if the index of $p$ in the integrand (or divergence $D$) is less than 0. Now, the propagator for the combined electroweak gauge bosons is

$$\Delta_{\mu\nu} = \frac{1}{p^2 - m^2} \left( -g_{\mu\nu} + (1 - \xi) \left( \frac{p_\mu p_\nu}{p^2 - \xi m^2} \right) \right),$$

where $\xi$ is the 't Hooft gauge term, which appears in the gauge fixing term in the Lagrangian for the interaction:

$$-\frac{1}{2\xi} (\partial_\mu A^\mu + \xi m \phi_2)^2 .$$

This term removes the unphysical (massless scalar) Goldstone boson $\phi_2$, which arises from the spontaneous symmetry breaking produced by the filled weak vacuum used to eliminate negative energy states. If $\xi$ is finite, then as $p_\mu \to \infty$, $\Delta_{\mu\nu} \to p^{-2}$, like the pure photon propagator, which, in the absence of any gauge choice, becomes:

$$\Delta_{\mu\nu} = \frac{1}{q^2} \left( -g_{\mu\nu} + (1 - \xi) \left( \frac{q^\mu q^\nu}{q^2} \right) \right).$$

However, for $\xi \to \infty$, we have the propagator for a massive vector boson theory without massless component,

$$\Delta_{\mu\nu} = \frac{1}{p^2 - m^2}\left(-g_{\mu\nu} + \left(\frac{p_\mu p_\nu}{m^2}\right)\right),$$

which becomes a constant when $p_\mu \to \infty$, leading to infinite sums in the diagrams equivalent to those in QED. One of the most convenient choices of gauge is $\xi = 1$ (Feynman gauge), which leads to an electroweak boson propagator,

$$\Delta_{\mu\nu} = \frac{-g_{\mu\nu}}{p^2 - m^2},$$

entirely analogous to that for the photon in the same gauge, and similarly linked by the factor ($kE + ii\mathbf{p} + ijm$) to the fermion propagator, as in QED. (It may be possible, here, to link the existence of massive weak bosons to the creation, in the nilpotent representation, of massive bosonic states via the interactions of fermions with the vacuum.)

The gluon propagator in QCD is again of the same form as that of the other gauge bosons:

$$\Delta_{\mu\nu} = \frac{1}{q^2}\left(-g_{\mu\nu} + (1-\xi)\left(\frac{q^\mu q^\nu}{q^2}\right)\right)\delta^{\alpha\beta},$$

which reduces, in Feynman gauge, to

$$\Delta_{\mu\nu} = \frac{1}{q^2}(-g_{\mu\nu})\,\delta^{\alpha\beta},$$

and which only differs from the electroweak boson propagator by a factor, $\delta^{\alpha\beta}$, arising from the fact that freely propagating gluons have fixed colours.

**10 Conclusion**

It is not necessary to follow in any further detail the procedures of QED, or derive further results by examining specific Feynman diagrams. The outline derivations in this paper show that the nilpotent method is not only amenable to QED calculations, but also has something additional to offer. It removes the infrared divergence from the integral for the fermion propagator by, in effect, combining the fermion and antfermion solutions in a single expression. By being already second quantized it requires the existence of finite values for fundamental quantities such as mass and charge at all energies, whatever integrals are performed to incorporate vacuum fluctuations, and the algebra itself provides successful methods for the automatic removal of all possible divergences. Since, the nilpotent formulation does not incorporate direct information about the *measure* of the coupling of charge to the field energy, then there is no fixed 'value' of charge to be renormalized, and perturbation expansions merely correlate the different measures of coupling which occur at different external field energies. Even when perturbation expansions need to be used, the theory removes divergences by *demanding* that there must be a finite upper limit to the energy used, which other work on the algebra tells us must be the Planck mass.[3,5] Because the process involved (via a filled vacuum) is the one which actually *creates* the masses of fermions, then our measure of the interaction of a charge with the field energy becomes simply an expected scaling mechanism, as happens also in classical contexts, rather than a desperate means of avoiding otherwise catastrophic results. It would appear that both the concept of 'renormalization' and the problem of divergences may be eliminated by using the nilpotent algebra, though the successful mathematical structure of QED remains intact.